\documentclass[revtex, amsmath, amssymb,superscriptaddress, reprint]{revtex4-2}

\usepackage[utf8]{inputenc}
\usepackage{graphicx}
\usepackage{dcolumn}
\usepackage{bm}
\usepackage[utf8]{inputenc}
\usepackage[T1]{fontenc}
\usepackage{mathptmx}
\usepackage{gensymb}
\usepackage{upgreek}
\usepackage{xcolor}
\usepackage{textcomp}
\usepackage{hyperref}
\usepackage{lineno}
\usepackage{mathtools,amssymb}
\usepackage{siunitx}

\hypersetup{colorlinks=true,linkcolor=red,citecolor=blue,filecolor=magenta,urlcolor=magenta}

\usepackage{titlesec}
\titleformat{\section}[hang]{\small\bfseries\sffamily}{\thesection.}{0.5em}{\MakeUppercase}
\titlespacing{\section}{0pc}{1pc}{0.2pc}
\usepackage{soul}
\begin{document}
\title{Controlling single rare earth ion emission in an electro-optical nanocavity}


\author{Likai Yang}
\affiliation{Department of Electrical Engineering, Yale University, New Haven, CT 06511, USA}
\author{Sihao Wang}
\affiliation{Department of Electrical Engineering, Yale University, New Haven, CT 06511, USA}
\author{Mohan Shen}
\affiliation{Department of Electrical Engineering, Yale University, New Haven, CT 06511, USA}
\author{Jiacheng Xie}
\affiliation{Department of Electrical Engineering, Yale University, New Haven, CT 06511, USA}
\author{Hong X. Tang}
\email{hong.tang@yale.edu}
\affiliation{Department of Electrical Engineering, Yale University, New Haven, CT 06511, USA}


\begin{abstract}
Rare earth emitters enable critical quantum resources including spin qubits, single photon sources, and quantum memories. Yet, probing of single ions remains challenging due to low emission rate of their intra-4$f$ optical transitions. One feasible approach is through Purcell enhanced emission in optical cavities. The ability to modulate cavity-ion coupling in real time will further elevate the capacity of such systems. Here, we demonstrate direct control of single ion emission by embedding erbium dopants in an electro-optically active photonic crystal cavity patterned from thin-film lithium niobate. Purcell factor over 170 enables single ion detection, which is verified by second-order autocorrelation measurement. Dynamic control of emission rate is realized by leveraging electro-optic tuning of resonance frequency. Using this feature, storage and retrieval of single ion excitation is further demonstrated, without perturbing the emission characteristics. These results promise new opportunities for controllable single photon sources and efficient spin-photon interfaces.

\end{abstract}
\maketitle

\section*{Introduction}

\noindent Quantum networks benefit from solid-state spins for their potential roles as quantum memories \cite{craiciu2019nanophotonic,de2008solid} and coherent transducers \cite{williamson2014magneto,bartholomew2020chip}. Addressing single atomic defect is also a key task in implementing solid-state qubit \cite{rose2018observation}, spin-photon entanglement \cite{togan2010quantum,bussieres2014quantum}, and single photon source \cite{rodiek2017experimental}. Among various systems studied, rare earth ion (REI) stands out thanks to its narrow, highly coherent transitions in both optical and microwave domains \cite{bottger2003material,serrano2018all}. In particular, erbium (Er) ion draws great attention due to its telecom-band emission. However, addressing weakly coupled $4f-4f$ optical transition of single REIs remains challenging. A critical solution to this is through Purcell enhancement of emission rate in optical cavities \cite{merkel2020coherent}. Using this method, optical probing of single REIs is recently achieved by imprinting micro-cavities onto bulk host crystals \cite{dibos2018atomic,zhong2018optically}. 

Toward better control and scalability of REI coupled systems, integrated photonic resonators with separate and instant tunability are highly desirable \cite{xia2022tunable,casabone2021dynamic}. Real-time modulation of emission rate could be used to design tailored single photon source. Such capacity is also advantageous in building high-efficiency spin-photon interface. A target transition can be switched on-resonance when strong drive is needed and switched off by detuning the cavity \cite{goswami2018theory}. It is also possible to modify spin cyclicity \cite{raha2020optical} by matching cavity frequency to different Zeeman transitions. As an example, branching ratio between optical transitions in an electronic spin $\Lambda$ system can be controlled by selectively enhancing the spin-conserving or spin-flipping transitions with the cavity. This could enable efficient spin pumping and initialization, which is vital in both quantum memory and qubit control applications \cite{kindem2020control,zhong2017nanophotonic}. Furthermore, individual tunablity will be particularly helpful when integrating systems with multi-qubit operation \cite{bernien2013heralded}.

Lithium niobate (LN) is a commonly used host crystal for REIs in both quantum and classical applications \cite{thiel2010optical,saglamyurek2011broadband,luo2021chip}. Meanwhile, it is also an important material in photonic technology with outstanding electro-optic \cite{xu2021bidirectional}, piezoelectric \cite{shen2020high}, and nonlinear \cite{lu2019periodically} properties. With recent development on thin film lithium niobate on insulator (LNOI), on-chip optical resonators such as rings \cite{zhang2017monolithic}, microdisks \cite{luo2017chip} and photonic crystals \cite{li2020lithium} are proposed and fabricated. Featuring high quality factor, small mode volume, and electro-optical tunability, these resonators are favorable platforms to access REIs. To date, incorporation of REIs into LNOI integrated photonics has been realized by ion implantation \cite{jiang2019rare,wang2020incorporation}, flip chip bonding \cite{yang2021photonic}, and smart-cut technique \cite{dutta2019integrated}. The last, in particular, guarantees full overlap between REIs and optical mode while preserving bulk coherence properties \cite{wang2022er}. Still, most of existing works focus on ion ensembles and single Er emitters in LN host are yet to be studied.

In this work, we demonstrate detection and control of single ion emission in smart-cut erbium doped lithium niobate (ErLN) thin film. Photonic crystal nanobeam cavities with electro-optical tunability are fabricated to achieve high Purcell enhancement of 177. By tuning the cavity to the tail of inhomogeneous broadening, single Er emission is spectrally resolved. It is then verified by second-order autocorrelation measurement yielding $g^2(0)=0.38$, indicating majority of collected photons come from a single ion.  We also show that the waveform of Er emission can be tailored using pulsed tuning voltage. The storage and retrieval of single ion excitation is further realized by detuning and re-aligning the cavity after the excitation pulse. The emission spectrum is shown not to be perturbed by this process. Our results pave the way for building tractable and scalable spin-photon interface with REIs.

\section*{Results}

\begin{figure*}
    \centering
    \includegraphics[width=1\textwidth]{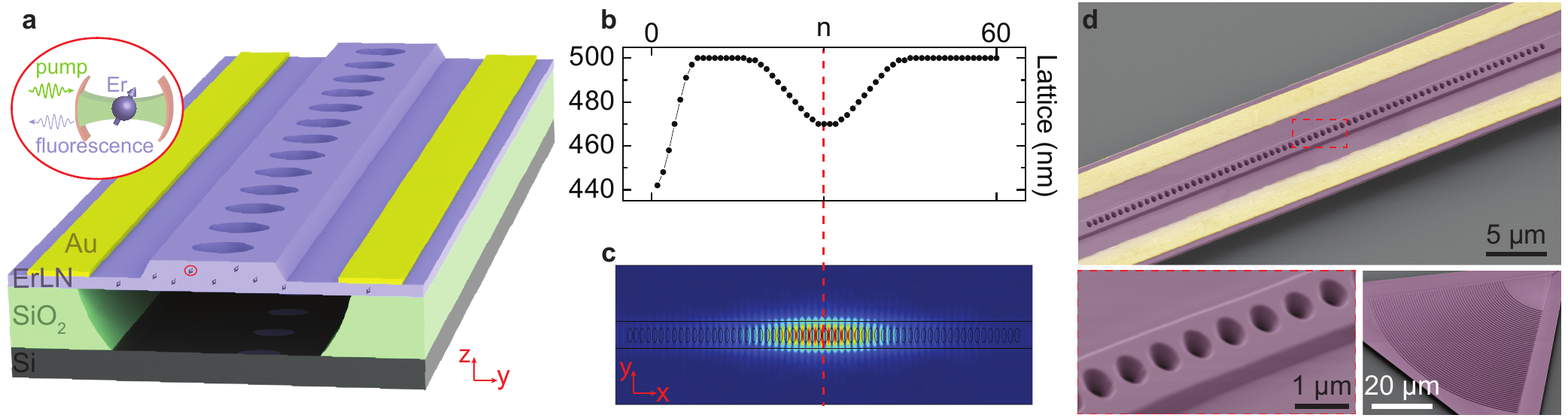}
    \caption{\textbf{Device design.} \textbf{a} Schematic drawing of our device (not to scale). Periodic holes are patterned on a suspended ridge waveguide to form photonic crystal mirror. Gold electrodes are deposited along the waveguide for electro-optic frequency tuning. Erbium (Er) ions are uniformly distributed in lithium niobate (LN). Ions in the photonic cavity can be addressed by collecting fluorescence. \textbf{b} Cavity design principle. The lattice constant is tapered down in the middle to support a defect mode and on one side for coupling. \textbf{c} Finite element simulation of resonance mode profile (fundamental TE). \textbf{d} False color SEM images of the actual device, with a zoom-in view on the waveguide. The grating coupler for fiber-to-chip coupling is also shown.}
    \label{fig1}
\end{figure*}

\noindent\textbf{Device design and preparation}. The devices are fabricated using 300\,nm z-cut ErLN (100\,ppm doped) thin films (from NanoLN), with a similar design concept previously used in developing LN electro-optic modulators \cite{li2020lithium}. Schematic drawing of the photonic crystal structure is shown in Fig.~\ref{fig1}a. On a half-etched ridge waveguide, periodic holes are patterned to form photonic crystal mirror. Silicon dioxide (SiO$_2$) layer beneath the waveguide is removed with buffered oxide etch (BOE) that goes through the holes. Such suspended structure improves mode confinement, thus is necessary for reaching high quality factor. Gold electrodes are then deposited along the waveguide for frequency tuning. Detailed fabrication process and device dimension is discussed in Supplementary Note 1. Er ions are uniformly distributed in LN and can be addressed via fluorescence emission. To form a cavity with a defect mode, lattice constant of the photonic crystal is tapered down in the middle, as plotted in Fig.~\ref{fig1}b. One side of the mirror is also tapered so that the cavity can be accessed by measuring the reflectivity. The simulated electric field profile of the optical fundamental TE mode is illustrated  in Fig.~\ref{fig1}c, with field component mostly in the crystal $y$-direction. With an in-plane voltage bias, the electro-optic coefficient $r_{22}=7\,\mathrm{pm/V}$ \cite{weis1985lithium} of LN is utilized. The half-etched slab helps to achieve larger tunability by increasing the overlap between optical mode and electric field. It also provides strong mechanical support to increase the fabrication yield. Fig.~\ref{fig1}d shows false color SEM images of the actual device, with a zoom-in view of the waveguide and the apodized grating coupler \cite{lomonte2021efficient} used for fiber-to-chip coupling.

To match Er emission wavelength, cavity resonance is swept in the fabrication process by varying the defect lattice constant. According to simulation, a 1\,nm difference in lattice constant will result in $\sim$3\,nm resonance shift. For our devices, we sweep across a 7\,nm range, sufficient to overcome uncertainty caused by fabrication imperfection and film thickness variation. Cryogenic electrical and optical access to the devices are realized by wire bonding and fiber glue techniques, respectively. The packaged devices are then loaded at 1\,K plate of a dilution refrigerator for cryogenic measurement.

\smallskip

\begin{figure}[!ht]
    \centering
    \includegraphics[width=0.49\textwidth]{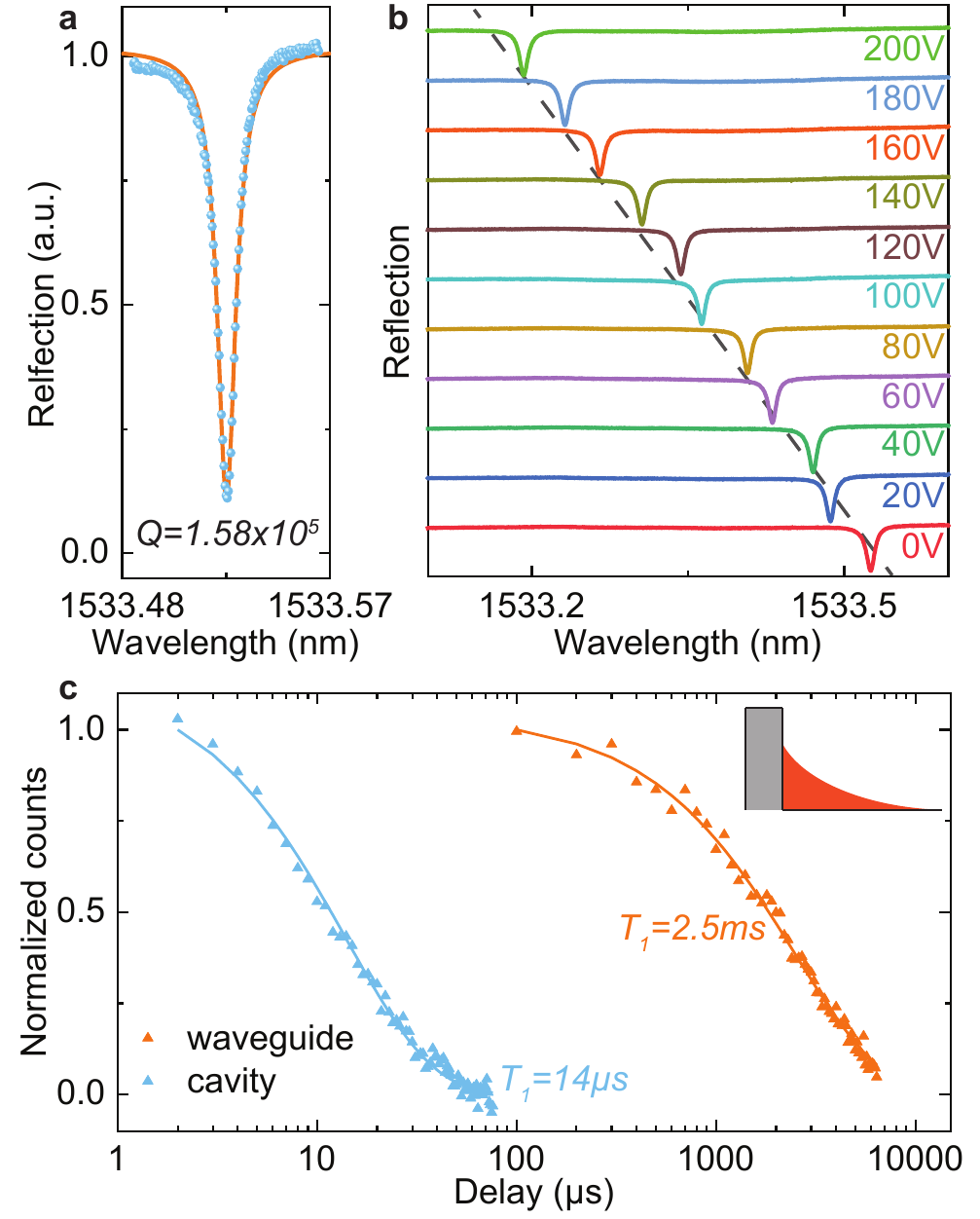}
    \caption{\textbf{Device characterization.} \textbf{a} Reflection spectrum of the resonator. A Lorentzian fit (orange curve) yields $Q=1.58\times10^5$. \textbf{b} Resonance frequency shift under different applied voltage. The reflection spectrum of each voltage is offset for clarity. A tuning rate of 1.6\,pm/V is extracted. \textbf{c} Time domain fluorescence measurement of erbium ions in the waveguide (orange) and cavity (blue) and exponential fitting. Enhancement of emission rate is shown by the shortened lifetime inside the cavity.}
    \label{fig2}
\end{figure}

\noindent\textbf{Device characterization}. At cryogenic temperature, the reflection spectrum of our device is first measured, as shown in Fig.~\ref{fig2}a. The resonance exhibits a quality factor of 158\,k and extinction ratio close to 10\,dB. The total throughput of our device is around 1\,\%. Electro-optic frequency tuning is calibrated in Fig.~\ref{fig2}b, by applying voltage up to 200\,V. Tuning rate of ~1.6\,pm/V is extracted. From our experience, breakdown of lithium niobate will usually be observed with a voltage larger than 500\,V. This gives a total tuning range of $\sim$1.5\,nm by applying bipolar voltage within this range.

\begin{figure*}
    \centering
    \includegraphics[width=1\textwidth]{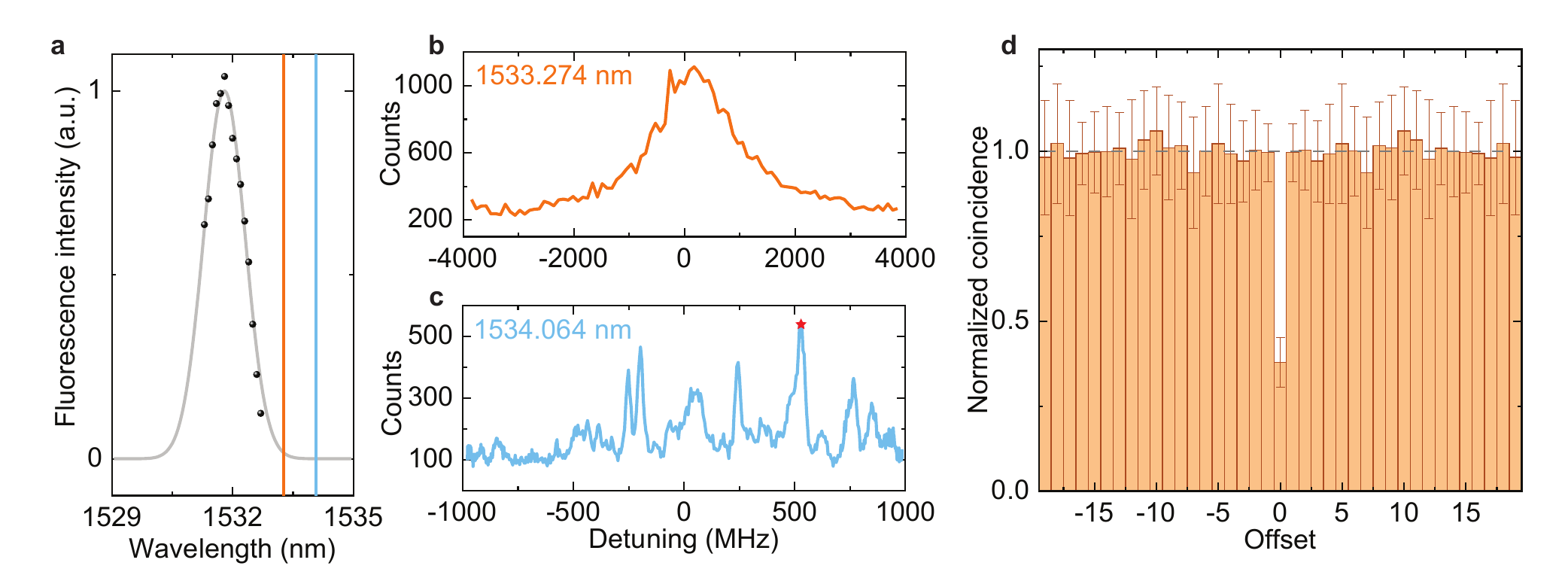}
    \caption{\textbf{Single ion detection.} \textbf{a} Inhomogeneous distribution of Er ions, measured by the fluorescence intensity in the waveguide. Gray curve is a Gaussian fit with FWHM of 160\,GHz. Orange and blue lines indicate the frequency that the cavity is tuned to for fluorescence spectrum measurement shown in panel b and c. \textbf{b} Fluorescence spectrum inside the resonance when it is tuned to 1533.274\,nm. A continuous emission spectrum shows that an ensemble of ions are probed. \textbf{c} Fluorescence spectrum when the resonance is tuned further to the tail of inhomogeneous broadening, at 1534.064\,nm. Discrete peaks suggest single ion emission. The star mark indicates the point for $g^{(2)}$ measurement. \textbf{d} Second-order autocorrelation function $g^{(2)}$ for the single ion emission. The $x$ axis is the number of offset between excitation pulses, $y$ axis is normalized coincidence. Calculated value of $g^{(2)}(0)=0.38\pm0.08$ proves that most of the collected photons come from a single ion. The data is symmetric around zero since a single detector is used.}
    \label{fig3}
\end{figure*}

Fluorescence from Er ions is collected using a commercial superconducting nanowire single photon detector (SNSPD). In the experiment, we use a pulse measurement setup in which the optical pump, SNSPD bias, and tuning voltage can all be controlled with synchronized pulse sequences (see Supplementary Note 2 for details). With a resonant fluorescence scheme, population lifetime of Er ions in the waveguide and the cavity are extracted, respectively. The results are plotted in Fig.~\ref{fig2}c. The cavity enhancement of Er emission is shown by reduction of lifetime from $T_{wg}\approx2.5$\,ms to $T_{cav}\approx14$\,$\mu$s. This yields a Purcell factor $P=T_{wg}/T_{cav}-1=177$. Theoretically, the Purcell enhancement of emission rate obeys the relation $P\propto Q/V_{mode}$ \cite{mcauslan2009strong}. Our device achieves strong enhancement by miniaturizing the mode volume $V_{mode}$ to $\sim$0.55\,$\mu$m$^3$ while maintaining high quality factor ($Q$). The measured average Purcell factor is also in good agreement with simulated value of 150 (Supplementary Note 3). 

\smallskip

\begin{figure*}
    \centering
    \includegraphics[width=1\textwidth]{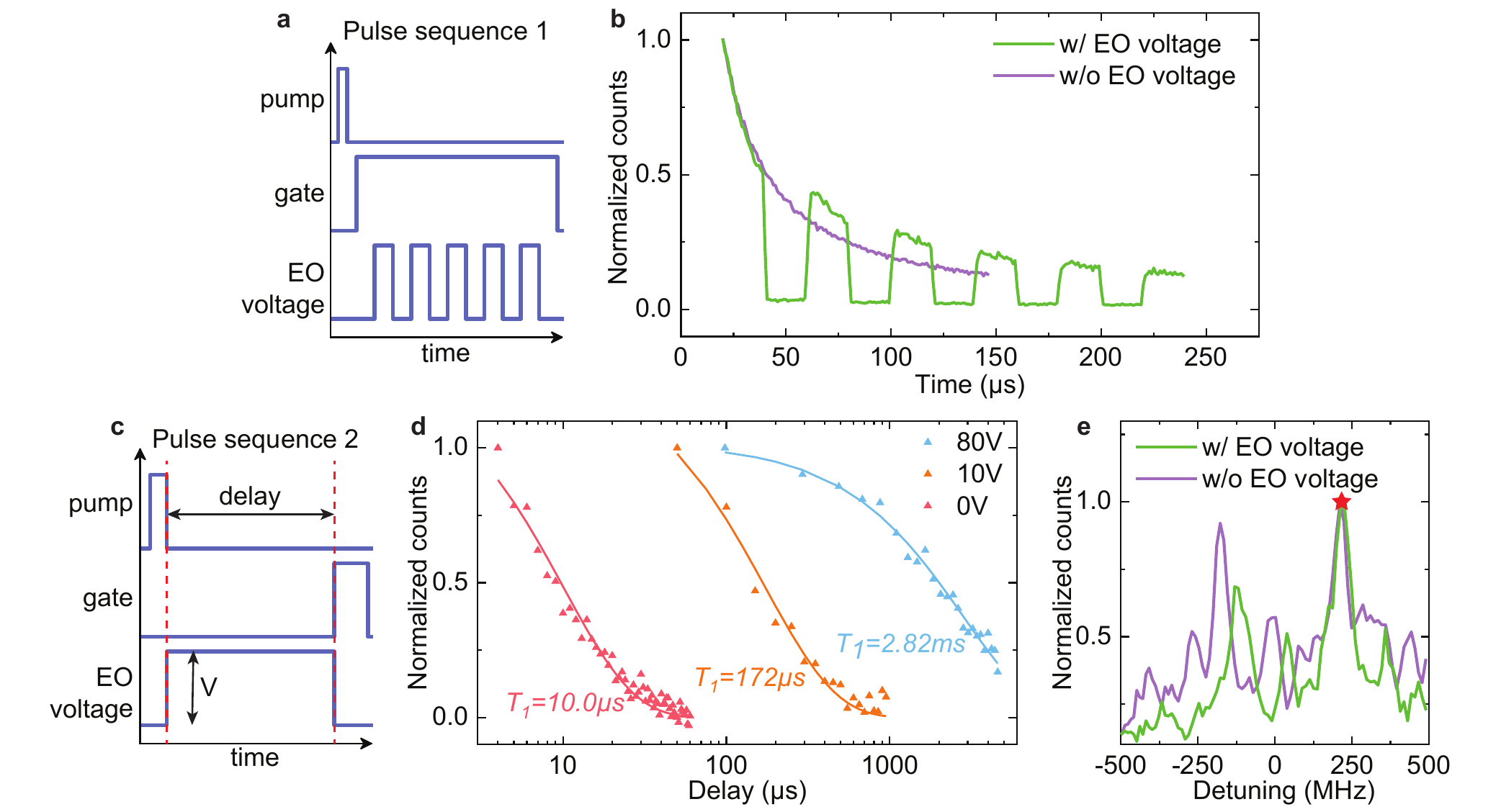}
    \caption{\textbf{Electro-optic control of Er emission.} \textbf{a} Pulse sequence used to demonstrate control of Er emission. Fluorescence signal is collected within a gate after the pump pulse. In the same time, the cavity frequency is modulated to suppress and enhance Er emission. \textbf{b} Measured results with pulse sequence in a. When electro-optic (EO) voltage is applied (green curve), a pulsed Er emission can be seen. The decay when cavity is on resonance corresponds well with the case without EO voltage (purple curve). \textbf{c} Pulse sequence for storage and retrieval of single ion excitation. After exciting a single ion, the cavity is detuned for a certain delay to store the excitation. The emission is then retrieved by tuning the cavity back on resonance. \textbf{d} Measured results with pulse sequence in c. The lifetime of single Er ion is first probed to be 10\,$\mu$s, shown by the red curve. The storage scheme is then applied with different tuning voltage. With 10\,V (orange) and 80\,V (blue), an extended lifetime of 172\,$\mu$s and 2.82\,ms is realized, respectively. \textbf{e} Fluorescence spectrum measured right after the pump pulse (purple curve) and after 100\,$\mu$s delay in the storage scheme (green curve). The single ion emission peaks match well for two cases. The small shift is potentially caused by laser drifting and spectral diffusion. This shows that the electro-optic tuning scheme does not perturb the emission spectrum. The red star mark indicates the peak used for measurement in c.}
    \label{fig4}
\end{figure*}

\noindent\textbf{Single ion detection.} Emission of Er $Y_1-Z_1$ transition in LN is inhomogeneously broadened, centering at $\sim$1532\,nm. Linewidth of the broadening can be measured by the fluorescence intensity in the waveguide at different wavelengths, as shown in Fig.~\ref{fig3}a. A full width at half maximum (FWHM) of 160\,GHz is extracted with a Gaussian fit, close to literature value of 180\,GHz. Electro-optic frequency tuning provides a convenient and deterministic way to park the cavity resonance at different positions of inhomogeneous distribution. For the following measurement, a device with $Q\approx100$\,k is used. We first tune the cavity to 1533.274\,nm and measure the spectrum of cavity enhanced fluorescence by sweeping the excitation laser across the resonance, as shown in Fig.~\ref{fig3}b. The corresponding fluorescence intensity shows a continuous peak that decays with the detuning, suggesting that an ensemble of ions are excited. In order to resolve single ion emission, we need to tune further away to the tail of inhomogeneous broadening, where the ions are more dilute spectrally. The results are shown in Fig.~\ref{fig3}c. The discrete peaks seen in the spectrum indicate that the fluorescence comes from single or several Er ions.

To confirm that the collected photons are from single ions, a second-order autocorrelation measurement is performed on one of the emission peaks (red star mark in Fig.~\ref{fig3}c). Since we are measuring resonant fluorescence signal with one detector, a time-bin analysis is imposed. In the experiment, fluorescence signal in a 10\,$\mu$s window after a 1\,$\mu$s excitation pulse is collected, with a repetition rate of 20\,$\mu$s. During data analysis, the photons collected after a single excitation pulse are pinned into one time bin. The coincidence events between time bins with offset $\tau$ are then extracted and normalized to calculate $g^{(2)}(\tau)$, while $g^{(2)}(0)$ corresponds to detecting two photons in a single time bin. The measured results are shown in Fig.~\ref{fig3}d, with $g^{(2)}(0)=0.38\pm0.08$. A simple signal-to-noise calculation of $g^{(2)}(0)$ (Supplementary Note 4) suggests that $\sim$80\,\% of the collected photons come from a single emitter. The remaining is attributed to SNSPD dark count ($\sim$10\,\%) and background emission ($\sim$10\,\%) from ions that are weakly coupled with the resonator, which are also the main factors that limit $g^{(2)}(0)$. This is in good agreement with a measured count rate of $\sim$200\,Hz when at the emission peak and $\sim$40\,Hz at the background. Another instability comes from the spectral diffusion of the ion and frequency drift of the laser which is not actively stabilized. We account for a total drift in the order of MHz/min in our system, which is compensated by manually tuning the laser during the $g^{(2)}$ measurement. Such drift also contributes to the linewidth of single ion emission peaks in Fig.~\ref{fig3}c, which ranges from 20 to 40\,MHz.

To further improve quantumness of the emission (i.e. smaller $g^{(2)}(0)$), a resonator with frequency further away from the 1532\,nm emission center can be used. This helps to suppress background emission by reducing the number of ions in the cavity. Also, the total transmission of our device is relatively low. With our design, transmission >10\,\% can be routinely achieved at room temperature, but is reduced during cooldown due to fiber misalignment caused by thermal contraction. Optimization of fiber gluing or improved grating coupler design can boost the collection efficiency of single ion emission. Frequency locking of the laser would also be helpful to extend the available probing time.

\smallskip

\noindent\textbf{Electro-optic control of Er emission.} Tuning the cavity in and out of resonance with the ions modifies the coupling between them, thus changes the emission rate. We demonstrate this by pulsing the tuning voltage while collecting the fluorescence. The corresponding pulse sequence used in the experiment is shown in Fig.~\ref{fig4}a. After exciting the ions on-resonance with the cavity, a 40\,V bias detunes the cavity so that Er emission into the cavity is suppressed, preserving the excited state population. The cavity can then be tuned back on resonance to restore ion-cavity coupling and enable the emission. This effect is clearly shown in Fig.~\ref{fig4}b, demonstrating real-time control of emission intensity. The switching time of this process is mostly limited by the amplifier used for generating high voltage, which has a bandwidth of around 1\,MHz.

With a similar protocol, storage and retrieval of single ion excitation can be achieved. To do so, we first focus onto a single ion emission peak. The lifetime of the ion is measured to be 10\,$\mu$s, as shown by the red trace in Fig.~\ref{fig4}d. A control sequence illustrated in Fig.~\ref{fig4}c is then applied. Here, the cavity is detuned right after pump pulses to store ion excitation. The fluorescence is then collected by tuning the cavity back after a certain delay. The storage lifetime of single ion excitation is extracted by varying the delay and the applied voltage, as shown in Fig.~\ref{fig4}d. When 10\,V bias is used, the cavity is not completely tuned away from the ion, so the lifetime extends to 172\,$\mu$s. Larger tuning voltage of 80\,V further recovers the ion lifetime to 2.82\,ms, similar to waveguide value. The slightly lengthened lifetime in detuned cavity could potentially be explained by the modification of spontaneous emission rate in photonic crystal band gap \cite{bayer2001inhibition,kaniber2007efficient}. The fluorescence spectrum after 100\,$\mu$s delay of cavity detuning is then measured and compared to the spectrum measured right after the pump pulse. The results are shown in Fig.~\ref{fig4}e. Similar peaks in these two curves suggest that the storage and retrieval process does not perturb the spectral property of the emission. The relative offset could come from the system drift discussed above. The red star mark indicates the point chosen for the storage measurement.

\section*{Discussion}
\noindent In conclusion, we have demonstrated the detection and electro-optic control of single Er emission in a LNOI photonic crystal nanobeam cavity. Exhibiting high quality factor and small mode volume, the resonator enhances the Er emission rate from a waveguide lifetime of 2.5\,ms to 14\,$\mu$s. Together with electro-optic tuning of resonance frequency to the tail of inhomogeneous broadening, single ion detection is enabled. A self intensity autocorrelation of $g^{(2)}(0)=0.38$ is measured, and practical ways to further improve single ion emission rate are discussed. The cavity is then shown to be able to store and shape single Er emission, with storage lifetime in detuned cavity matching the waveguide value. Utilizing electric field along $y$-direction, tuning rate of 1.6\,pm/V is achieved. Larger tunability could be realized with a different geometry or film orientation. For example, the electro-optic coefficient $r_{33}$ along $z$-direction is five times larger than the $r_{22}$ we use here. The advantage of using $r_{22}$ is that the DC stark shift of ErLN vanishes in this orientation \cite{wong2002properties}, so that it could be separated from electro-optic tuning. Combining with the ability to control ion frequency by Zeeman effect, full control over REI-cavity coupling system can be envisioned. The tunablity is also particularly useful when multiple REIs are to be addressed in a single cavity for frequency multiplexing, or integrating REIs coupled with different cavities on a single chip.

Although the decoherence and spectral diffusion property seen in ErLN is not as good as certain other hosts such as yttrium orthosilicate (YSO), it can be greatly improved by applying external magnetic field. Our previous work measured coherence time as long as 180\,$\mu$s on smart-cut ErLN, under 0.5\,T magnetic field and milli-kelvin temperature. This is already much longer than the cavity enhanced lifetime in our resonator. Using a lower doping concentration will also help to improve coherence by suppressing Er-Er interaction. Along with the capability to integrate on-chip microwave resonators, efficient manipulation of single rare earth spin is highly feasible based on our device. Another interesting perspective lies in the strong piezoelectricity of LN, as the effect of strain on REIs has been previously demonstrated \cite{zhang2020inhomogeneous}. The coupling between single REIs and mechanical mode \cite{molmer2016dispersive} can be studied. With all these features, our device provides solid foundation for versatile and efficient REI based spin-photon systems.

~\\

\section*{Data availability}

\noindent The datasets that support the plots within this paper and other findings of this study are available from the corresponding author (H.X.T.) upon reasonable request.

\section*{Acknowledgement}

\noindent This work is supported by the US Department of Energy Co-design Center for Quantum Advantage (C2QA) under contract No. DE-SC0012704. We acknowledge initial funding from the Department of Energy, Office of Basic Energy Sciences, Division of Materials Sciences and Engineering under Grant DE-SC0019406. H.X.T. acknowledges partial support from the National Science Foundation (NSF) through ERC Center for Quantum Networks (CQN) grant EEC-1941583. The authors acknowledge helpful discussion with Dr. Charles W. Thiel and Dr. Baptiste Royer. The authors would like to thank Dr. Yong Sun, Kelly Woods, and Dr. Michael Rooks for their assistance provided in the device fabrication. The fabrication of the devices was done at the Yale School of Engineering \& Applied Science (SEAS) Cleanroom and the Yale Institute for Nanoscience and Quantum Engineering (YINQE).

\section*{Author contributions} 

\noindent H.X.T. and  L.Y. conceived the experiment. L.Y. designed and fabricated the device. L.Y., S.W., M.S., and J.X. contributed to the preparation of the device. L.Y. and S.W. performed the experiment. L.Y. wrote the manuscript with contribution from all authors. H.X.T. supervised the work.

\section*{Competing interests} 

\noindent The authors declare no competing interests.


\end{document}


\title{Supplementary Information for "Controlling single rare earth ion emission in an electro-optical nanocavity"
}


\author{Likia Yang}
\affiliation{Department of Electrical Engineering, Yale University, New Haven, CT 06511, USA}
\author{Sihao Wang}
\affiliation{Department of Electrical Engineering, Yale University, New Haven, CT 06511, USA}
\author{Mohan Shen}
\affiliation{Department of Electrical Engineering, Yale University, New Haven, CT 06511, USA}
\author{Jiacheng Xie}
\affiliation{Department of Electrical Engineering, Yale University, New Haven, CT 06511, USA}
\author{Hong X. Tang}
\email{hong.tang@yale.edu}
\affiliation{Department of Electrical Engineering, Yale University, New Haven, CT 06511, USA}



\maketitle

\newpage
\section*{Supplementary note 1. device fabrication}
\begin{figure}[h]
    \centering
    \includegraphics[width=1\textwidth]{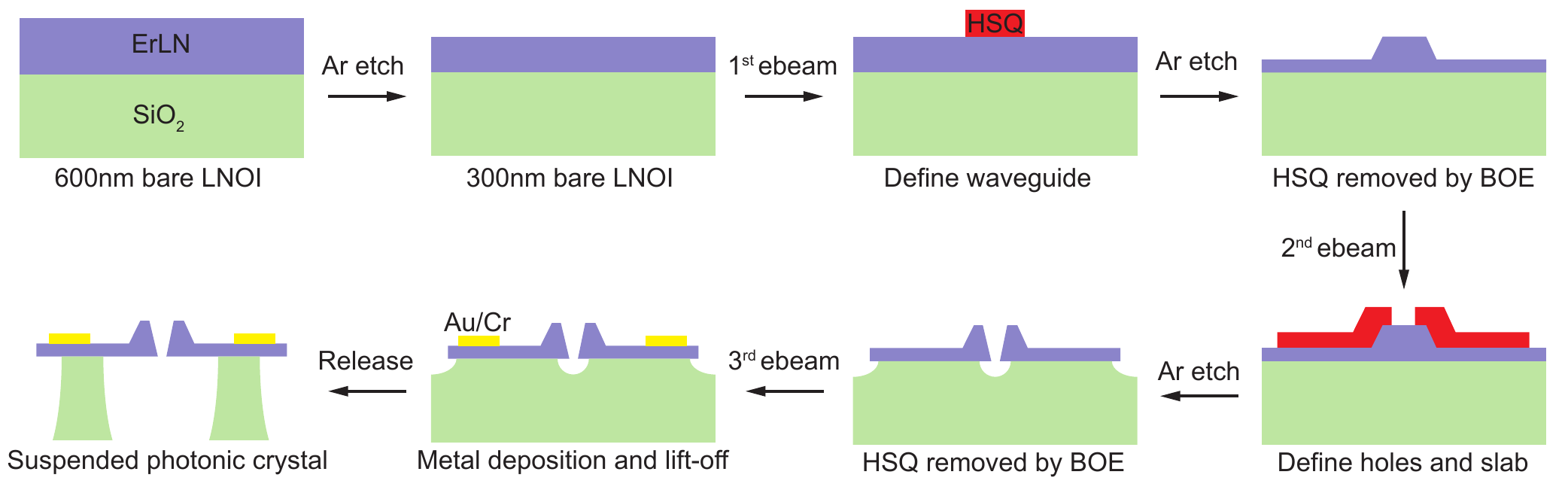}
    \caption{\textbf{Fabrication process for photonic crystal devices.}}
    \label{figS1}
\end{figure}

The fabrication process of our devices is described in Fig.~\ref{figS1}. We start with a 600\,nm erbium (Er) doped lithium niobate on insulator (LNOI) film, with 2\,$\mu$m silicon dioxide (SiO$_2$) as substrate. It is first thinned down to 300\,nm by a reactive ion etching (RIE) process with argon (Ar) plasma. The waveguide is then defined by electron beam lithography (EBL), using hydrogen silsesquioxane (HSQ) as resist. Another Ar plasma RIE etches 180\,nm into LN to form a ridge waveguide. The waveguide width is set to be 1.2\,$\mu$m and the angle of etching process is $\sim$60$\degree$. After etching, the residue resist is removed by buffered oxide etch (BOE). The second EBL defines photonic crystal holes and the slab with HSQ. The hole dimensions are 600\,nm$\times$350\,nm. The EBL dose is carefully adjusted so that the actual dimensions of the holes match well with designed values. After that, a RIE process etches through LN. Removal of residue resist with BOE will cause small undercut in the SiO$_2$ layer, but the device structure is robust enough for further fabrication. The third EBL uses polymethyl methacrylate (PMMA) to lift off metal electrodes, which consist of 5\,nm chromium (Cr) under 50\,nm gold (Au) deposited by thermal evaporation. The thin Cr layer is used to improve the adhesion. The gap between metal electrodes is designed to be 5\,$\mu$m so that the optical mode will not be perturbed. We do not see noticeable difference in optical quality factor with or without the metal electrodes.  Finally, the chip is dipped into BOE for a longer time to form a suspended structure.

\section*{Supplementary note 2. Measurement setup}

\begin{figure}[h]
    \centering
    \includegraphics[width=1\textwidth]{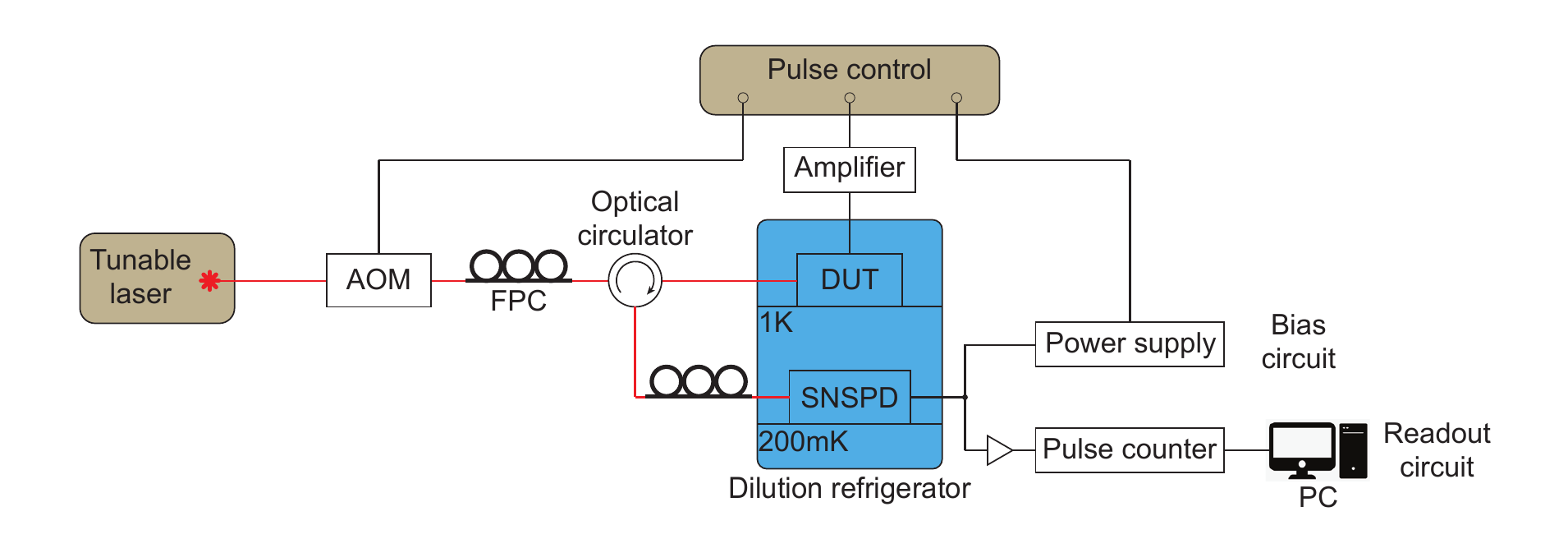}
    \caption{\textbf{Schematic drawing of measurement setup.} AOM: acousto-optic modulator. FPC: fiber polarization controller. DUT: device under test. SNSPD: superconducting nanowire single photon detector.}
    \label{figS2}
\end{figure}

The schematic drawing of our measurement setup is shown in Fig.~\ref{figS2}. Light from a tunable laser (Santec TSL-710) is chopped by acousto-optic modulators (AOM) to generate excitation pulses. In the case of frequency sweeping, the internal piezoelectric tuning function of the laser is utilized. On-off extinction ratio of >100\,dB is reached using 2 AOMs. The light is then sent to our device under test (DUT) mounted at 1\,K plate of a dilution refrigerator. The reflection as well as fluorescence signal from DUT is collected via an optical circulator and sent to a superconducting nanowire single photon detector (SNSPD), sitting at 200\,mK of the same fridge. Pulsed electro-optic tuning voltage is generated by a high voltage amplifier. The amplifier has a bipolar maximal output voltage of 200V and bandwidth $\sim$1\,MHz. The bias circuit of the SNSPD consists of a pulse generator and an in-line 50\,k$\Omega$ resistor. This allows the SNSPD to be turned off during excitation pulse to prevent saturation, and turned on in the fluorescence collection window. Readout of SNSPD signal is done by a pulse counter connected to PC. The pulse counter (Picoharp 300) works in a real-time data collection mode, in which the time tag of each detected photons can be registered. The AOM, tuning voltage, and SNSPD bias pulses are all synchronized with controlled delay and width.

\section*{supplementary note 3. Purcell enhancement}
Theoretically, the Purcell enhancement of an Er ion inside a cavity can be expressed as \cite{miyazono2017coupling}
\begin{equation}
    P(\vec{r})=\frac{3}{4\pi^2} \frac{\beta Q \lambda^3}{\chi_L n^3V_{mode}} \frac{|E(\vec{r})|^2}{|max(E(\vec{r}))|^2},
\end{equation}
where $\beta=0.22$ \cite{mcauslan2009strong} is the branching ratio of radiative transition of ErLN, $n\approx2$ is the cavity refractive index, and $\chi_L=[(n^2+2)/3]^2\approx2$ is the local field correction. The mode volume is defined as $V_{mode}=\frac{\int \epsilon|E(\vec{r})|^2d\vec{r}}{max(\epsilon |E(\vec{r})|^2)}$, with integral over all space. The Purcell enhancement of an ion depends on the field strength $|E(\vec{r})|$ at the ion location. The average Purcell enhancement of the cavity is then the average over all ions, weighted by their field intensity:
\begin{equation}
    P_{avg}=\frac{\int_{LN} P(\vec{r})|E(\vec{r})|^2d\vec{r}}{\int_{LN}|E(\vec{r})|^2d\vec{r}}=\frac{3}{4\pi^2} \frac{\beta Q \lambda^3}{\chi_L n^3V_{eff}}.
\end{equation}
Here, the effective mode volume $V_{eff}$ can be expressed as
\begin{equation}
    V_{eff}=\frac{\int|E(\vec{r})|^2d\vec{r}\int_{LN}|E(\vec{r})|^2d\vec{r}}{\int_{LN}|E(\vec{r})|^4d\vec{r}}.
\end{equation}
This can be calculated from finite element simulation of the cavity mode profile, yielding $V_{eff}=2\,\mu$m$^3$. This gives $P_{avg}=150$.

\begin{figure}[h]
    \centering
    \includegraphics[width=1\textwidth]{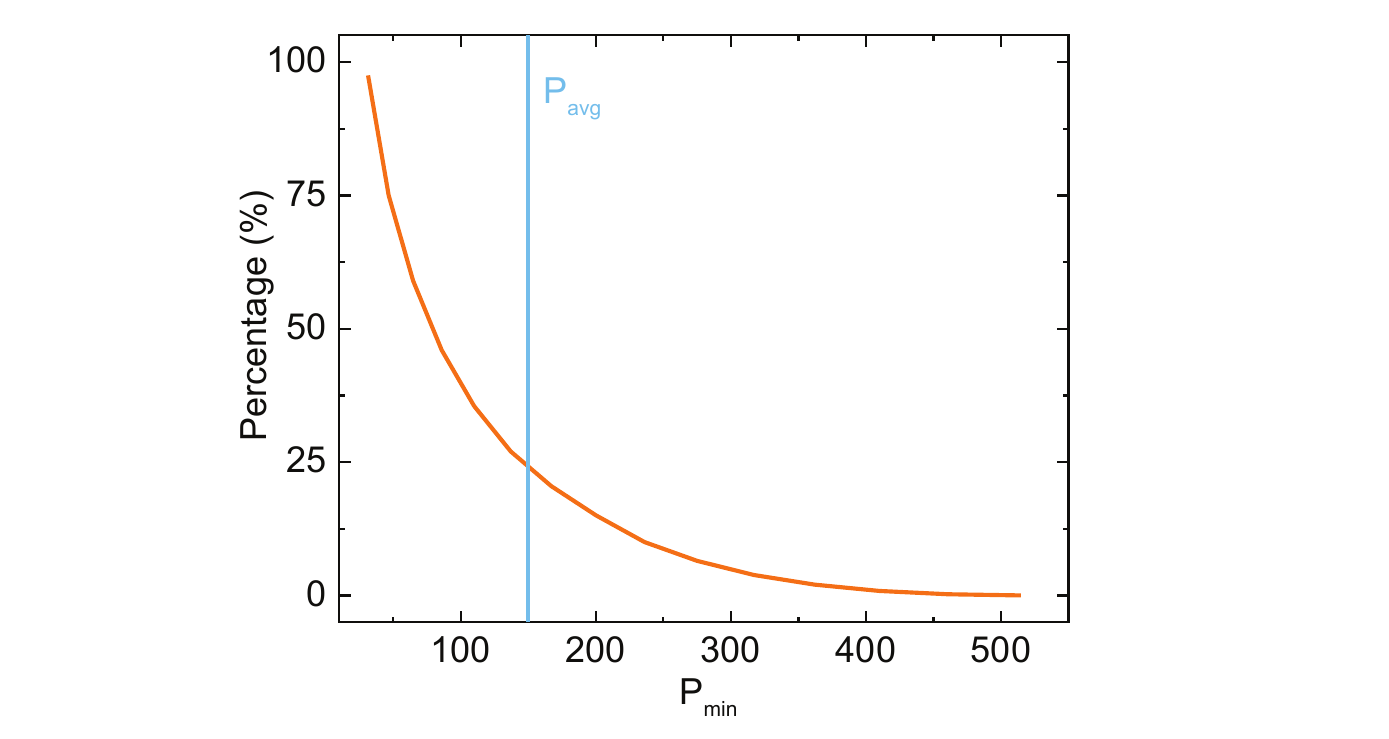}
    \caption{\textbf{Distribution of Purcell enhancement for ions in the cavity.} The $y$ axis denotes the percentage of ions that has Purcell factor larger than $P_{min}$, with respect to all ions in a volume $V_{eff}=2\,\mu$m$^3$.}
    \label{figS3}
\end{figure}

The distribution of Purcell factor for ions in the cavity can also be extracted using these equations. We calculate the percentage of ions in the cavity that has Purcell factor larger than different $P_{min}$. The results are shown in Fig.~\ref{figS3}.

\section*{supplementary note 4. single ion count rate and $g^{(2)}$ analysis}
Theoretically, the count rate we get from a single Er ion in the cavity can be calculated as \begin{equation}
    N=P_e \times P_{decay} \times P_{cav-wg} \times P_{fiber-chip} \times P_{loss} \times \eta_{SNSPD} \times \frac{1}{T_{rep}}.
    \label{eqS1}
\end{equation}
Here, $P_e=1/2$ is the maximum probability an ion being in excited state after an incoherent pump. $P_{decay}=1-e^{t/T_1}$ is the probability it decays to ground state during the collection window $t$. For us, $t=10\,\mu$s and $T_1=10\,\mu$s, so $P_{decay}=0.63$. $P_{cav-wg}=\frac{\kappa_{ex}}{\kappa_{ex}+\kappa_{in}}\approx 1/2$ is the coupling rate between cavity and waveguide. $P_{fiber-chip}=0.1$ is the single side fiber-to-chip coupling efficiency. Loss in optical components such as the circulator and the fibers are included in $P_{loss}\approx 0.6$. Collection efficiency of our SNSPD is $\eta_{SNSPD}=50$\,\%. These account for the number of photons detected after each single excitation pulses. It is then multiplied by the repetition rate $1/T_{rep}=50$\,kHz to get the actual count rate. The resulting value is 236\,Hz, in rough agreement with the $\sim$160\,Hz we get from our measurement.

The photons we collect for second-order correlation measurement are from two parts, the single ion $I_{ion}$ and the background $I_{bg}$. The background is attributed to the emission from ions weakly coupled to the cavity and the dark count from SNSPD. Here, we take $g^{(2)}_{ion}(0)=\langle I_{ion}^2\rangle/\langle I_{ion}\rangle^2=0$ and $g^{(2)}_{bg}(0)=\langle I_{bg}^2\rangle/\langle I_{bg}\rangle^2=1$. The signal-to-noise ratio is defined as $SNR=\langle I_{ion}\rangle/\langle I_{bg}\rangle$. Then, the measured second-order autocorrelation function would be
\begin{equation}
    g^{(2)}(0)=\frac{\langle I^2\rangle}{\langle I\rangle^2}=\frac{\langle (I_{ion}+I_{bg})^2 \rangle}{\langle I_{ion}+I_{bg}\rangle^2}=\frac{\langle I_{ion}^2\rangle+\langle I_{bg}^2\rangle+2\langle I_{ion}\rangle\langle I_{bg}\rangle}{\langle I_{ion}\rangle^2+\langle I_{bg}\rangle^2+2\langle I_{ion}\rangle\langle I_{bg}\rangle}=\frac{2SNR+1}{(SNR+1)^2}.
\end{equation}
For us, we get $g^{(2)}(0)=0.38$. This gives $SNR=3.70$, suggesting that $\sim$79\,\% of the collected photons are from a single ion. This is in agreement with our estimation of $\sim$160\,Hz single ion count rate and $\sim$40\,Hz background count rate.

Apart from device optimization to increase Purcell enhancement, improvement of $g^{(2)}(0)$ can mainly come from two aspects. The first is improving fiber-to-chip coupling efficiency. With optimized fiber glue process and better grating coupler design, a single-side transmission of 50\,\% is achievable. This will increase single ion count rate to $\sim$1000\,Hz. The other factor lies in suppression of background Er emission. This can be done by using smaller doping concentration or tuning the cavity frequency further away. The ultimate background is the dark count from SNSPD, which is $\sim$20\,Hz in our case. Implementing these improvements will result in $g^{(2)}(0)\approx 0.04$.


\bibliographystyle{naturemag}
\bibliography{ref_si}